\documentstyle[12pt]{article}

\begin{document}
\setbox1 = \hbox{{{$\lambda$}}}
\def\R{{{$\lambda$}}\hskip-\wd1{\raise2pt\hbox{$-$}}\hskip\wd1}
\def\lambar{\hbox{{\R}}}
\begin{titlepage}
\hoffset=.5truecm
\voffset=-2truecm

\centering

\null
\vskip 1truecm
{\normalsize \sf \bf International Atomic Energy Agency \\
and \\
United Nations Educational,Scientific and Cultural Organization\\}
\vskip 1truecm
{\huge \bf
INTERNATIONAL CENTRE \\
FOR \\
THEORETICAL PHYSICS\\}
\vskip 3truecm


{\LARGE\bf
Precocious scaling in antiproton--proton scattering\\
 at low energies}\\
\vskip 1truecm

{\large \bf
{ D.B.Ion
\\}
\normalsize Institute of Atomic Physics,P.O.Box MG-6,Bucharest,
{\bf Romania \\ }
\medskip
{\large \bf V.Topor Pop\\}
\normalsize International Centre for Theoretical Physics,Trieste 34100,
 {\bf Italy\\}
\normalsize and\\}
\normalsize INFN Sezione di Padova,University of Padova,Via Marzolo 8,
Padova,{\bf Italy\\}
\medskip
{\large \bf C.Petrascu \\}
\normalsize Institute for Atomic Physics,P.O.Box MG - 6,Bucharest,
{\bf Romania\\}
{\normalsize and \\}
{\large \bf V.Popa \\}
\normalsize Institute for Space Sciences,P.O.Box MG - 6,Bucharest,
{\bf Romania}

\vskip 2truecm
{\it Work presented at International Conference Particles and
Nuclei PANIC XIII--Perugia,June 28 -July 2 1993,Italy}

\vskip 8truecm

\end{titlepage}

\hoffset = -1truecm
\voffset = -2truecm

\title{\bf Precocious scaling in antiproton--proton scattering\\
at low energies}
\author{
{\bf D.B.Ion
}\\
\normalsize Institute of Atomic Physics,P.O.Box MG-6,Bucharest,
{\bf Romania}\\
{\bf V.Topor Pop}\\
\normalsize International Centre for Theoretical Physics,Trieste 34100,
{\bf Italy}\\
{\normalsize and}\\
\normalsize INFN Sezione di Padova,University of Padova,Via Marzolo 8,
Padova,{\bf Italy}\thanks{On leave from absence from Institute for Space
Sciences,P.O.Box MG - 6,Bucharest,
{\bf Romania} E-MAIL TOPOR@ROIFA.BITNET(EARN)}\\
{\bf C.Petrascu}\\
\normalsize Institute of Atomic Physics,P.O.Box MG-6,Bucharest,
{\bf Romania}\\
{\normalsize and}\\
{\bf V.Popa}\\
\normalsize Institute for Space Sciences,P.O.Box MG - 6,Bucharest,
{\bf Romania}
}
\date{29 July 1993}
\newpage
\maketitle

\begin{abstract}

The scaling of the diffraction peak in antiproton--proton
scattering has been investigated from near threshold up to
$3\,\,\, GeV/c$ laboratory momenta.It was shown that the scaling
of the differential cross sections are evidentiated with a
surprising accuracy not only at high energies,but also at very
low ones(e.q. $p_{LAB}=0.1--0.5\,\,\, GeV/c$),beyond the resonance and
exotic resonance regions.This precociuos scaling strongly suggests
that the s -- channel helicity conservation (SCHC) can be a peculiar
property that should be tested in antiproton -proton interaction
not only at high energies but also at low energy even below  $ p_{LAB}
 = 1\,\,\, GeV/c $.

\end{abstract}

\newpage


\section{Introduction}

The description of the elementary particles scattering in
terms of optimal state  (minimum norm principle)
was recently presented in refs [1-5].Then two important physical
laws for the hadron--hadron scattering (the scaling of the
angular distribution and s--channel helicity conservation (SCHC))
were derived, using Reproducing Kernel Hilbert Space
(RKHS) methods.

Starting with the assumption that each of the two bodies
helicity scattering amplitude is an element of RKHS,it was
shown that (see refs.[2-5]):

i)  The RKHS has many special properties that make it
an adequate space for the description of the scattering
in terms of a minimum norm principle;

ii)  The notion of optimal scattering state[1,3,6,7] and
reproducing Kernel(RK) of the RKHS - associated to the
helicity amplitude,are the same;

iii)  The expansion of the scattering amplitudes in terms
of optimal states can be important alternative to the
phase shift analysis;

iv)  The dual diffractive scattering (DDS) [8,9] and the
dual diffractive resonance phenomena (DDR) [6,7,8] are
described in an unified manner using RKHS methods.

v)  The scaling of the angular distributions ,as well as
the s-channel helicity conservation (SCHC) in hadron-hadron
scattering,are two important properties of the optimal
states.

 In this paper we present an optimal state analysis (OSA)
on the $\bar p p$ scattering at low energies(laboratory
momenta between 181 MeV/c and 3000 MeV/c) from the point
of view of the scaling property for differential cross
sections.We have chosen this energy range for the
interest since it containe the energy region for production
of the resonances or exotic resonances (multiquarc states)
in $ \bar p p$ scattering .A great experimental efforts
(Chrystal Barrel,Obelix,E760 Fermilab,GAMS,JETSET,
BIS--2 ,DAFNE)and theoretical ones are done for search of
exotic multiquark resonances. For some reviews see
references [10--14].

The possibility to find criteria for identify resonance
signals , via violation of the scaling property, is also one
of the motivation of this paper.

\section{Optimal State Analysis in hadron-hadron scattering}

    Let us consider the two body elastic scattering  :
\begin{equation}
a  + b \rightarrow a + b
\end{equation}

where particle a and b have spins $ s_a $ and $ s_b $.

In describing system (1) the helicity formalism of Jacob
and Wick [15] was used.

Therefore, let:
\begin{equation}
 f^{[\mu]}(x) = \langle\mu_{a}^{'}\mu_{b}^{'}\vert
\,F(s,t) \,\vert\mu_{a}\mu_{b}\rangle
\end{equation}
be the helicity amplitudes of the system (1);where $ \mu_{a} $
and $ \mu_{b} $ are the initial helicities and $ \mu_{a}^{'} $
and $ \mu_{b}^{'} $ the final ones ; s and t are the usual
Mandelstam variable ,while $ x = \cos\theta$,$ \theta$ beeing
the c.m. scattering angle.The normalization was chosen in
such a way that the differential and elastic cross section for
the channel \(  [\mu] = (\mu_{a},\mu_{b};\mu_{a}^{`},\mu_{b}^{`}) \)
are given by :
\begin{equation}
\frac{d\sigma^{[\mu]}}{d\Omega}(x)  =
\vert f^{[\mu]}(x)\vert^{2}\,\,\,\,\,\,x \in [-1,+1]
\end{equation}
\medskip
\begin{equation}
\sigma_{el}^{[\mu]} = 2\pi \int\limits_{-1}^{+1}\vert f^{[\mu]}(x)
\vert^{2}dx = 2\pi\Vert f^{[\mu]}\Vert^{2}
\end{equation}

Since we will work at fixed energy the dependence of
$ f^{[\mu]}(x)$ ;$\frac{d\sigma^[\mu]}{d\Omega}(x)$;
$\sigma_{el}^{[\mu]}$;$\sigma_{el}$ and $\sigma_{T}$
of this variable is supressed.

Let $ \frac{d\sigma}{d\Omega}(x)$; $\sigma_{el}$ and  $\sigma_{T}$ be
the unpolarized (differential ,elastic and total) cross sections:
\begin{equation}
\frac{d\sigma}{d\Omega}(x)=\frac{1}{(2s_a+1)(2s_b+1)}=
\sum_{[\mu]}\,\,\frac{d\sigma^{[\mu]}}{d\Omega}(x)
\end{equation}
\begin{equation}
\sigma_{el}=\frac{1}{(2s_a+1)(2s_b+1)}\sum_{[\mu]}\,\,\sigma_{el}
^{[\mu]}
\end{equation}
\begin{equation}
\sigma_{T}=\frac{1}{(2s_a+1)(2s_b+1)}\sum_{[\mu_{o}]}\,\,
\sigma_{T}^{[\mu_{o}]}
\end{equation}

where $ [\mu_{o}]= (\mu_{a},\mu_{b};\mu_{a},\mu_{b})$ denote
helicity conserving channels.

A step toward  the description of the scattering in terms of
an optimum principle (minimum norm principle) is to consider each
amplitude $ f^{[\mu]}(x)$ as an element of a functional RKHS
[4,5],$H^{[\mu]}$ defined on $ S=[-1,+1]$,with inner product and
norm given by :
\begin{equation}
\langle f^{[\mu]}(x), g^{[\mu]}(x) \rangle =
\int\limits_{-1}^{+1} f^{[\mu]}(x)\bar g^{[\mu]}(x)\,dx
\end{equation}
\medskip
\begin{equation}
\Vert f^{[\mu]}\Vert^{2} = \langle f^{[\mu]}(x), g^{[\mu]}(x)\rangle
< \infty
\end{equation}

If $K^{[\mu]}$ is the reproducing kernel of $ H^{[\mu]}$,
$ f^{[\mu]} \in  H^{[\mu]}$ then for each $y\in [-1,+1]$
for which $f^{[\mu]}(y)\ne 0 $,$K^{[\mu]}(y,y)\ne 0 $ the
functional (3) and (4) obey the inequality :
\begin{equation}
\frac{{d\sigma}^{[\mu]}}{d\Omega}\le
\frac{\sigma_{el}^{[\mu]}}{2\pi}\,\,K^{[\mu]}(y,y)
\end{equation}
The equality holds in (10) if and only if :
\begin{equation}
f^{[\mu]}(x)= f^{[\mu]}(y)\frac{K^{[\mu]}(x,y)}{K^{[\mu]}(y,y)}
\,\,\,\,\,\,x\in[-1,+1]
\end{equation}

The scattering state of the system (1) described by the helicity
amplitude (11) is called optimal state (OS) of the channel $[\mu]$.
Writing the scattering amplitude $f^{[\mu]}$ in terms of the
partials amplitudes $f^{[\mu]}_j$,using the rotation functions
$d^{j}_{\mu \nu}(x)$ as the orthonormal sequence of RKHS
$H^{[\mu]}$ ,we obtain:
\begin{equation}
f^{[\mu]}(x) = \sum_{j_{min}}^{j_{\mu}}(2j+1)
f_j^{[\mu]}d_{\mu \nu}^{j}(x)
\end{equation}

$$j_{min}=max\{|\mu|,|\nu|\}\,\,\,\,\,\,\mu=\mu_a-\mu_b,
\nu=\mu_a^{'} - \mu_b^{'}$$

Now,one can prove that [2-5]:

a)the helicity amplitude $f^{[\mu]}$ is an element of a
RKHS , $H^{[\mu]}$,defined on $[-1,+1]$,if and only if
$j_{\mu} < \infty$

b)$H^{[\mu]}$ possesses the following reproducing kernel:
\begin{equation}
K^{[\mu]}(x,y) = \sum_{j_{min}}^{j_{\mu}}(j+1/2)
d_{\mu \nu}^{j}(x)d_{\mu \nu}^{j}\,(y)
\end{equation}
\begin{equation}
K^{[\mu]}(x,y) = \frac{[(j_{\mu}+1)^{2} - \mu^{2}]^{1/2}
[(j_{\mu}+1)^{2} - \nu^{2}]^{1/2}}{2(j_{\mu}+1)}\times
\frac{d_{\mu \nu}^{j_{\mu}+1}(x)d_{\mu \nu}^{j_{\mu}}(y)
-d_{\mu \nu}^{j_{\mu}}(x)d_{\mu \nu}^{j_{\mu}+1}(y)}{x - y}
\end{equation}
\begin{equation}
K^{[\mu]}(y,y) = \frac{[(j_{\mu}+1)^{2} - \mu^{2}]^{1/2}
[(j_{\mu}+1)^{2} - \nu^{2}]^{1/2}}{2(j_{\mu}+1)}\times
\{ \dot d_{\mu \nu}^{j_{\mu}+1}(y)d_{\mu \nu}^{j_{\mu}}(y)-
\dot d_{\mu \nu}^{j_{\mu}}(y)d_{\mu \nu}^{j_{\mu}+1}(y) \}
\end{equation}
where
$$ \dot d_{\mu \nu}^{j}(x) = d\,\,d_{\mu \nu}^{j}(x)/dx$$

Next,let us assume that each helicity amplitude ,$f^{[\mu]}$,is
an element of a RKHS $ H^{[\mu]}$,which possesses the RK
given by equations (13-15). Then, the following important
results are obtained (see ref.[4,5]):

i) If $ \sigma_{el}$ and  $\frac{d\sigma}{d\Omega}(1)$ are given,
any cut-off $j_{\mu}$ on the total angular momentum must
obey the bound :
\begin{equation}
(j_{\mu}+1)^{2} \ge \frac{4\pi}{\sigma_{el}}
\frac{d\sigma}{d\Omega}(1) + \mu^{2}
\end{equation}

ii) the equality is obtained in (16) if and only if
$f^{[\mu]}(x)$ is the optimal amplitude :
\begin{equation}
f^{[\mu]}(x)= f^{[\mu]}(1)\frac{K^{[\mu]}(x,1)}{K^{[\mu]}(1,1)}
\end{equation}
\medskip
\begin{equation}
f^{[\mu]}(x)= f^{[\mu]}(1)\frac{1}{j_{\mu}+1}
\frac{d_{\mu \mu}^{j_{\mu}+1}(x)-d_{\mu \mu}^{j_{\mu}}(x)}{x-1}
\end{equation}

where

\begin{equation}
j_{\mu}=integer\left(half-integer\right)
\{ [\frac{4\pi}{\sigma_{el}}
\frac{d\sigma}{d\Omega}(1) + \mu^{2}]^{1/2} -1 \}
\end{equation}

iii) the logaritmic slope of the forward diffractive peak is
independent of $ [\mu] $ and is given by :
\begin{equation}
b_{o}=\frac{d}{dt}\left[\ln \frac{d\sigma}{d\Omega}(s,t)\right]_{t=0}=
b_{o}^{[\mu]} = \frac{{\lambar}^{2}}{4}[\frac{4\pi}{\sigma_{el}}
\frac{d\sigma}{d\Omega}(1) - 1]
\end{equation}
with $\lambar = 1/p_{cm}$ and  $p_{cm}$ are the momenta in c.m system.

iv) The forward diffraction peak of the optimal phenomena,described
by equations (16-20),possesses the scaling property
(function $ R_{th}(\tau) $)
\begin{equation}
\frac{d\sigma}{d\Omega}(x)/
\frac{d\sigma}{d\Omega}(1)= [\frac{2J_{1}(\tau)}{\tau}]^{2}
\end{equation}
with the scaling variable
\begin{equation}
\tau = 2 \{ |t|b_{o} \}^{1/2}
\end{equation}
\medskip
\begin{equation}
\tau = \{\lambar^{2}|t|[\frac{4\pi}{\sigma_{el}}
\frac{d\sigma}{d\Omega}(1) - 1]\}^{1/2}
\end{equation}
where $J_{1}(\tau)$ is the Bessel,first order,function.

We remark that the results from relations (22) and (23)
include in a more general and exact form the scaling
variables $|t|\sigma_{T}^{2}/4 \pi \sigma_{el}$ and
$|t| \sigma_{T}$ introduced by Sing and Roy [16],
Cornille and Martin [17],Dias de Deus [18] and
Buras and Dias de Deus [19] .

\section{Numerical results on $\bar p p $ elastic scattering
at low energies(precocious scaling)}
In a recent paper [20] high energy data on $ \bar p p $
elastic scattering were analysed,via a single optimal
state analysis (SOSA) and it was shown that,at high energies
$ p_{LAB} \geq 3\,\,\, GeV/c  $ the optimal predictions are satisfied
experimentally to a good accuracy.

Also a detailed comparison of the optimal state predictions with
the experimental data for the elastic slope was performed [1,20].

Now,we present  some results for $ \bar p p $ elastic
scattering data analysis at low energies obtained via SOSA,
which have been described briefly in Section 2.
The laboratory  momenta of antiprotons range between 180 MeV/c
and 3000 MeV/c ,including the region of resonances and exotic
resonances.We chosed this interesting energy range in order to
find possible ways of identifying signals given by resonances,
which can manifest in SOSA analysis by scaling  (and SCHC)
violations.

Therefore,using the optimal state predictions(see equations 21-23)
we have computed the following quantities:

a) The scaling variable $ \tau $ given by equation (23)
(The experimental data  was taken from
reference [21],[22],[33--36])

b) The experimental scaling function
$$
R_{exp}(\tau) =\frac{d\sigma}{d\Omega}(x)/
\frac{d\sigma}{d\Omega}(1)
$$

c) The theoretical scaling function $ R_{th}(\tau)$ given by
ecuation (21).

The numerical results obtained in this way are given in
figures 1[a--d];2[a--f];3[a--d];4[a--d];5[a--d].
Analysing these figures we can see that the scaling of
the differential cross section is a property evidentiated
with a "surprising accuracy ",outside the resonance
region.(See figures 1[a--d],2[a--f],3[a--d] and 4[a--d].
A violation of this scaling ( even at lowest values of
$ \tau $ is appearing clearly in the resonances and exotic
resonances region (see figures 5[a--d]).

 The discrepancy,at higher values of scaling variable
$ (\tau \geq 2.5) $ (see figures 4[a--d]) could be avoided,
especialy for the second peak region ,if two or more
optimal state will be included in the theoretical calculations.
However,in the resonance region the discrepancy between
theoretical SOSA prediction and experimental data is very
large. In order to check some possibilities to identify
specific properties of resonances we performed a more
detailed analysis for the incident momenta between 180 MeV/c
up to 3000 MeV/c(experimental data taken from references
[21--36].So,a statistical analysis of chi squared
distributions $( \chi^{2}/NDF) $ in a 3--dimensional plot
for all values of scaling variable $ \tau $ (see figure 6)
and for scaling variable $ \tau \leq 3 $ (Figure 7) are
presented.

Also ,bidimensional plots for all values of scaling variables
(Figure 8) and for values of $ \tau \leq 3 $ (Figure 9) are done.
We note that no selection in the data was chosen.
Some characteristic signals for particular values of scaling
variable are clearly seen in these figures ,but more
rafined analysis are required ,in order to derive some
definite conclusions.

\section{Conclusions}

{}From the analysis presented in this paper (see Section 3) we
see that a good agreement between SOSA-predictions and the
experimental data is obtained only in the range  beyond
the resonance  region.The discrepances  between SOSA predictions
and experimental scaling function appear only in the resonance
region(see figures 5[a--d]and figures 6--9).It is important
to note that the presented results are only preliminary OSA
analysis in this energy domain .The hint is that rafining
our analysis on large sample of data,will be able to derive
some criteria to evidentiate the resonance signals and
part of theirs properties.

   Finally,we note that we intend to perform also an analysis with
two ( or more ) optimal states .This is required not only by the
discrepancies observed (see Figures 4[a--d] ,but also for taking
into account the spin polarisation phenomena inhadron--hadron
scattering .

\section{Acknowledgements}

The authors of this paper(V.T.P ,V.P and C.P) would like  to  express
their gratitude to Professor Abdus Salam and to International Atomic
Energy Agency  and UNESCO for hospitality at the International
Centre for Theoretical Physics ,Trieste where this paper was completed.

We would like to thank to directors of International Scools
 "Non Accelerator Particle Physics ",Professor G.Giacommelli,
Professor N.Paver,and to Professor R.Daemi and Professor  F.Hussain
for kind invitation and finacial  support during the
"Summer School in High Energy and Cosmology",ICTP - Trieste.

One of us (Vasile Topor Pop) acknowledge  financial support from
INFN - Sezione di Padova,where part of this work was done.

\newpage

\begin{thebibliography}{99}

\bibitem[1]{kn:ion1}D.B.Ion,Rev.Roum.Phys.{\bf 37}1,(1992)
\bibitem[2]{kn:ion2}D.B.Ion and H.Scutaru,Int.Journ.of Theor.
Phys.{\bf 24},355(1985)
\bibitem[3]{kn:ion3}D.B.Ion,Rev.Roum.Phys.{\bf 36},251(1991)
\bibitem[4]{kn:ion4}D.B.Ion,Int.Journ.of Theor.
Phys.{\bf 25},1257(1986)
\bibitem[5]{kn:ion5}D.B.Ion,Int.Journ.of Theor.
Phys.{\bf 24},1217(1985)
\bibitem[6]{kn:ion6}D.B.Ion,R.Ion Mihai,Nucl.Phys.{\bf A360},
400(1981)
\bibitem[7]{kn:ion7}D.B.Ion,R.Ion Mihai,Rev.Roum.Phys.{\bf 36},
15(1991)
\bibitem[8]{kn:ion8}D.B.Ion,Rev.Roum.Phys.{\bf 26},15(1981)
\bibitem[9]{kn:ion9}D.B.Ion,Rev.Roum.Phys.{\bf 26},25(1981)
\bibitem[10]{kn:lan}{\it Surveys in High Energy Physics}1992
{\bf 6} pp 257--414 (Harwood Academic Publishers GmbH,Printed
in United Kingdom)
\bibitem[11]{kn:kle}E.Klempt,Sov.J.Nucl.Phys.{\bf 55},942(1992),
Proceedings of the Workshop on Nucleon-Antinucleon Interactions,
ITEP,Moscow,8--11,July 1991
\bibitem[12]{kn:mon}L.Montanet,Proceedings of the 3rd International
Conference on Hadron Spectroscopy,HADRON 89 ,Ajaccio,Corsica
(France)September 23--27(1989)(Eds.F.~Binon,J.~M.~Frere,
J.~P.~Peigneux)pp 668(1989)
\bibitem[13]{kn:gua}G.Guaraldo ,{\it Lectures delivered at the
International NATO ASI Advanced School,Il Ciocco,Italy,
June 1992}(to be published in Proceedings)
\bibitem[14]{kn:sil}B.Silvestre Brac,C.Semay,
Z.Phys.{\bf C57}273(1993)
\bibitem[15]{kn:jac}M.Jacob and G.C.Wick,Ann.of Phys.(N.Y),
{\bf 7},404(1959)
\bibitem[16]{kn:sin}V.Sing,S.M.Roy,Phys.Rev.{\bf D1},2638(1970);
Phys.Lett.{\bf 24},28(1970)
\bibitem[17]{kn:cor}H.Cornille,A.Martin, CERN - Report ,
TH--2130(1976)
\bibitem[18]{kn:dia1}J.Dias de Deus,Nucl.Phys.{\bf 159},231(1973)
\bibitem[19]{kn:dia1}J.Dias de Deus,Nucl.Phys.{\bf 375},981(1974)
\bibitem[20]{kn:ion10}D.B.Ion,C.Petrascu,Rev.Roum.
Phys.{\bf 37},569(1992)
\bibitem[21]{kn:comp} $\bar N$N and $\bar N$D interactions,
A Compilation Particle Data Group (Eds James E.Enstrom et al.),
L.B.L-58(1972)p 83
\bibitem[22]{kn:bru}W.Brucner et al.,Z.Phys.{\bf A339},367(1991)
\bibitem[23]{kn:sp}D.Spencer and N.D.Eduards,Nucl.Phys.
{\bf B19},501(1970)
\bibitem[24]{kn:cork}B.Cork et al.,Nuovo Cimento{\bf 25},497(1962)
\bibitem[25]{kn:conf}B.Conforto et al.,Nuovo Cimento,{\bf 54A},
441(1968)
\bibitem[26]{kn:clin}D.Cline et al.,Phys.Rev.Lett.{\bf 21},1268,
(1968)
\bibitem[27]{kn:kalb}G.R.Kalbfleisch,R.C.Strand and V.Vanderburg
,Nucl.Phys.{\bf B30},466, (1971)
\bibitem[28]{kn:baco}T.C.Bacon et al.,Nucl.Phys.{\bf B32},66(1971)
\bibitem[29]{kn:park}D.L.Parker et al.,Nucl.Phys.{\bf B32},29(1971)
\bibitem[30]{kn:cooper}W.A.Cooper et al.Nucl.Phys.{\bf B16},155(1970)
\bibitem[31]{kn:dom}V.Domingo et al.,Phys.Lett.{\bf 24B},642(1967)
\bibitem[32]{kn:esco}D.Escoubes et al.,Phys.Lett.{\bf 5},
132(1963)
\bibitem[33]{kn:kunn}R.A.Kunne et al.,Nucl.Phys.{\bf B323},1(1989)
\bibitem[34]{kn:scia}P.Sciavon et al.,Nucl.Phys.{\bf A505},
595(1989)
\bibitem[35]{kn:lins}L.Linssen et al.,Nucl.Phys.{\bf A469},
726(1987)
\bibitem[36]{kn:bose}P.Bosetti et al.,Nucl.Phys.{\bf B94},
21(1975)

\newpage

\begin{figure}
\label{Figure 1 [a--d].}
\caption{Optimal state predictions are compared with experimental
data from reference [22]}
\label{Figure 2 [a--f]}
\caption{Optimal state predictions are compared with experimental
data from reference [25]}
\label{Figure 3 [a--d]}
\caption{Optimal state predictions are compared with experimental
data from reference [23]}
\label{Figure 4 [a--d]}
\caption{Optimal state predictions are compared with experimental
data from reference [27]}
\label{Figure 5 [a--d]}
\caption{Optimal state predictions are compared with experimental
data from reference [29]}
\label{Figure 6}
\caption{3--dimensional plot for all values of scaling
 variable $ \tau $.Experimental data are from references
[22--36]}
\label{Figure 7}
\caption{3--dimensional plot for scaling variable $ \tau \leq 3 $.
Experimental data are from references [22--36]}
\label{Figure 8}
\caption{bi--dimensional plot for all values of scaling
 variable $ \tau $.Experimental data are from references
[22--36]}
\label{Figure 9}
\caption{bi--dimensional plot for scaling variable $ \tau \leq 3 $.
Experimental data are from references [22--36]}
\end{figure}

\end{document}